\documentstyle[12pt]{article}
\newcommand{\beq}{\begin{equation}}
\newcommand{\eeq}{\end{equation}}
\newcommand{\bea}{\begin{eqnarray}}
\newcommand{\eea}{\end{eqnarray}}

\begin{document}
\thispagestyle{empty}
\vspace*{-15mm}
\baselineskip 10pt
\begin{flushright}
\begin{tabular}{l}
{\bf OCHA-PP-195}\\
{October (2002)}\\
{\bf hep-ph/0210235}
\end{tabular}
\end{flushright}
\baselineskip 24pt 
\vglue 10mm 
\begin{center}
{\LARGE\bf 
Generation of 4d Gauge Theory and Gravity from their 3d versions\footnote{Talk given at GRG11 international conference held at Tomsk, Russia on July 1-6 (2002)} 

--Asymptotic Disappearance of Space and Time (ADST) Scenario--
}
\vspace{7mm}

\baselineskip 18pt 
{\bf
Akio SUGAMOTO}
\vspace{2mm}

{\it 
 Department of Physics, Ochanomizu University, \\
Tokyo 112-8610, Japan
}\\
\vspace{10mm}
\end{center}

\begin{center}
{\bf Abstract}\\[7mm]
\begin{minipage}{14cm}
\baselineskip 16pt
\noindent

An exotic scenario of our universe is proposed in which our universe starts from zero space-time dimensions (0d), namely, a set of discrete points, it increases the (continuous) dimensionality during the cooling down of it, and finally arrives at the present four space-time dimensions (4d).  This scenario may be called asymptotic disappearance (or discretization) of space-time (ADST) scenario.  The final stage of the scenario is to generate dynamically 4d from the 3d, which is shown to be possible for both gauge theory and gravity, including the standard model.  
To examine the validity of the scenario, 4d QED generated from its 3d version at high energy is studied, in which one spatial dimension is discretized with a lattice constant $a$.  From the LEP2 experiment on $e^+ e^- \to \gamma \gamma$,  $a$ is constrained to satisfy $a \le 461$ GeV.  Expected bound on $a$ at future $e^+ e^-$ linear collider is discussed. 
The finite size of $a$ modifies the dispersion relation, causing the violation of Lorentz symmetry. Then, the paradox of observing $20$ TeV $\gamma$ rays from the active galaxy Markarian 501 may  disappear, overcoming the GZK cut.

\end{minipage}
\end{center}


\newpage
\baselineskip 18pt 
\def\thefootnote{\fnsymbol{footnote}}
\setcounter{footnote}{0}

\section{Usual scenario v.s. ADST scenario}
 
I am very happy to give a talk at this GRG11 international conference on theoretical and  experimental problems of general relativity and gravitaion held in Tomsk, on the occasion of celebrating 100 anniverasry of  Tomsk State Pedagogical University.  My another job of coming here is to make the international exchange program to start between Tomsk State Pedagogical University and our Ochanomizu University in Tokyo.  Now, I will come back to my first job.

In the usual scenario, our universe  starts with 10 dimensions (10d), described probably by the 10d superstring theories. During the cooling down of the universe, the space-time dimensions are reduced from 10d to 4d at present low energy universe.  Contrary to this usual scenario, our universe starts from 0d (a set of discrete points) in our scenario, and the process of cooling down increases the space-time dimensions from 0d to 4d one by one.  If this scenario is viewed from low energy to high eneregy, the space-time dimensions disappear asymptotically.  Therefore, we may call our scenario "asymptotic disappearnce of space and time (ADST) scenario".  It may be better to use "discretization" instead of "disappearnce", if we consider what happens really.  The ADST scenario is motivated by the paper of Arkani-Hamed, Cohen and Georgi~\cite{Georgi} in which 5d gauge theory is generated from 4d gauge theory at high energy.  They have called the mechanism "(De) Construction".  

We discuss mainly on the final stage of the ADST scenario in which our present 4d theories are generated from their 3d versions  at high energy.  My talk is based on our three works~\cite{Sugamoto}, \cite{CIS}, and~\cite{CSW},  performed in collaboration with Gi-Chol Cho, Etsuko Izumi, and Isamu Watanabe.

\section{Gauge theory}
Let's start with N copies of D dimensional gauge theory at high energy. Each gauge theory is located on the D dimensional world sheets (branes), labeled by $n=1-N$.  Let the gauge group be $G=SU(m)$, and we may call it "flavor" gauge group. Our starting action is 
\beq
S=-\frac{1}{2 (g_{D})^{2}}  \int d^{D}x \sum_{n=1}^{N} tr \left( F_{ij}(x,
n) F^{ij}(x, n) \right). 
\label{3D gauge action}
\eeq
We assume that the other $(n+\frac{1}{2})$-th brane is prepared between $n$-th and $(n+1)$-th branes, and on this $(n+\frac{1}{2})$-th brane the strong gauge theory with gauge group $G_s=SU(n_s)$ works.  We may call this second gauge group "color" gauge group.  If the energy decreases, the "color" singlet pair condensates $U(n, n+1)$ of "mesons"  can be formed.  For this to occur we have to introduce "quark" $\chi(n, n+\frac{1}{2})$ and "anti-quark" $\chi(n+\frac{1}{2}, n+1)$.  Then, the "meson" is formed as
\beq
U(x; n, n+1) = \frac{1}{2\pi (f_{s})^{D-1}} 
\langle \chi(n, n+\frac{1}{2})\chi(n+\frac{1}{2}, n+1)\rangle,
\label{condensation}
\eeq
where $f_{s}$ is the decay constant of "meson". 
A characteritic feature of this fermions~\cite{Georgi} is that the "quark" has "flavor" on the $n$-th brane, and the "anti-quark" has "anti-flavor" on the $(n+1)$-th brane, having "color" on the $(n+\frac{1}{2})$-th brane.  Then, the generated "meson", $U(n, n+1)$, belongs to the fundamental and the anti-fundamental representations of "flavor" gauge group on the $n$-th and the $(n+1)$-th branes, respectively, so that $U(n, n+1)$ becomes a link variable.  It can be written as  $e^{iaA_{0}(x, n)}$ with  $A_{0}(x, n)$, the gauge field of a newly generated dimension, and $a$, its lattice constant.

Fluctuation of the condensates gives the generated action at low energy:
\beq
\Delta S = (f_{s})^{D-2}~\int d^{D}x ~\sum^{N}_{n=1} tr
\left[\left(D_{i}U(x; n, n+1)\right)^{\dagger}\left(D^{i}U(x; n,
n+1)\right)\right].
\eeq
Now, we can easily understand that the sum of $S+\Delta S $ is the "flavor" gauge theory in (D+1) dimensions in the lowest order approximation in $a$.  For this to occur, the generated diminsion is preferably space-like with the following relations:
\beq
a=\frac{1}{g_{D}(f_{s})^{(D-2)/2}} ~~~\mbox{and}
~~~(g_{D+1})^{2}=a(g_{D})^{2}. 
\label{conditons}
\eeq

\section{Gravity}
We start with N copies of 3d gravity having $SO(4)$ (or $SO(1, 3)$) gauge group:
\begin{equation}
S_{G} =\frac{1}{2 (\kappa_{3})^{2}}\sum_{n=1}^{N}\int d^3x
\frac{1}{2}\epsilon^{ijk} B^{AB}_{i}(x, n)R^{AB}_{jk}(x, n).
\label{original 3D gravity action}
\end{equation}
Here, $B^{AB}_{i}(x, n)$ is the  $SO(4)$ gauge fields and 
$R^{AB}_{jk}$ is the $SO(4)$ curvature.  Since $SO(4)=SU(2) \times SU(2)$, the $B$ fields are decomposed into two sets of 3d dreibeins
\begin{equation}
B^{ab}_{i}=\frac{1}{2} \epsilon_{abc}(e^{c}_{i}+\bar{e}^{c}_{i}),~~
B^{0a}_{i}=\frac{1}{2} (-e^{a}_{i}+\bar{e}^{a}_{i}).
\end{equation}
Then, the action becomes the sum of two sets of 3d Einstein gravities: 
\begin{equation}
S_{G} =\frac{1}{2(\kappa_{3})^{2}} \sum_{n=1}^{N} 
\int d^3x \frac{1}{2} \epsilon^{ijk} 
\left(e^{a}_{i}(n)R^{a}_{jk}(n)
+\bar{e}^{a}_{i}(n)\bar{R}^{a}_{jk}(n) \right). 
\label{3D gravity}
\end{equation}
As pointed out by Witten~\cite{Witten}, this is a kind of topological field theory, and is less singular than the 4d Einstein gravity.

If we can set
\begin{equation}
B^{AB}_{i}
=\frac{1}{2} \epsilon^{ABCD}E_{C0}E_{Di},
\label{condition}
\end{equation}
in terms of 4d vierbeins, then the actin becomes a part of the 4d Einstein gravity. The original $B^{AB}_{i}$ have eighteen degrees of freedom, but vierbein $E^{C}_{\mu}$ have sixteen degrees of freedom.  If we impose six natural constraints, that is the two 3d metrics are identical,
\beq
g_{ij}=\bar{g}_{ij}, ~~\mbox{for}~~
g_{ij}=\sum_{c} e^{c}_{i}e^{c}_{j} ~~\mbox{and}~~ \bar{g}_{ij}=\sum_{c}
\bar{e}^{c}_{i}\bar{e}^{c}_{j},
\label{eqaul metrics}
\eeq
or equivalently impose,
\beq
\frac12 \epsilon^{ABCD} B^{AB}_{i} B^{CD}_{j}=0,
\eeq
then we can express $B^{AB}_{i}$ in terms of $E^{C}_{\mu}$.  Here, four degrees of freedom in $E^{C}_{\mu}$ are redundant, but these degrees of freedom, say $E^{C}_{0}$, are exactly the necessary degrees of freedom in 4d, since the distance between the adjacent branes can be locally fluctuated.

Similarly as before, if the energy decreases, the fermion condendates can appear, due to the extra strong gauge theories:
\begin{equation}
\frac{1}{2\pi f_{g}^{2}} \langle 
\lambda(n,n+\frac12)\lambda(n+\frac12, n+1) \rangle 
= O(x;n, n+1)=e^{a\omega_{0}(x, n)},
\end{equation}
and the new components of curvature appear as
\beq
D_i O(x, n, n+1)^{AB}   
=a R^{AB}_{i0}(x, n) + \cdots. 
\eeq
Now, the missing part of the 4d gravity is recovered by
\begin{equation}
\Delta S_{G} = -f_{g}^{2} \int d^{3}x ~a\sum_{n=1}^{N} 
\frac{1}{4}\epsilon^{ijk}\epsilon_{ABCD}E^{A}_{i}(x, n)
R^{CD}_{k0}E^{B}_{j}(x, n+1) +\cdots,
\end{equation}
and $S_{G}+\Delta S_{G}$ becomes 4d Einstein gravity at low energy. 
Of course, the following conditions are required,
\begin{equation}
a=
\frac{1}{2(\kappa_{3})^{2}(f_{g})^{2}}~\mbox{and}~(\kappa_{4})^{2}
=2a(\kappa_{3})^{2}.
\end{equation}
 
Generalization of my work to arbitrary D dimensions is done by Myron Bander~\cite{Bander}.
He starts with N copies of the $SO(1, D)$ invariant generalized Dd gravity,
\bea
S_{G}=\sum^{N}_{n=1}&&M^{D-2}_{(D+1)} \int d^{D}x~~ \epsilon^{i_1, \cdots, i_{D}} \epsilon_{A_0, \cdots, A_{D}} \nonumber \\
&& \times E^{A_0}_0 E^{A_1}_{i_1} \cdots E^{A_{D-2}}_{i_{D-2}} R^{A_{D-1}A_{D}}_{i_{D-1}i_{D}},
\eea
generates the following action 
\bea
\Delta S_{G}=\sum^{N}_{n=1} &&2M^{D-1}_{(D+1)} \int d^{D}x~~ \epsilon^{i_1, \cdots, i_{D}} \epsilon_{A_0, \cdots, A_{D}} \nonumber \\
& & \times E^{A_0}_{i_1} E^{A_1}_{i_2} \cdots E^{A_{D-2}}_{i_{D-1}} R^{A_{D-1}A_{D}}_{i_{D}, 0} \times a.
\eea
and obtain the (D+1)d Einstein gravity.

\section{Standard model}

The gauge interactions differ in the standard model for L- and R- fermions, namely, $SU(2)_{L} \times U(1)_{Y}$ and $U(1)_{Y}$, so that the gauge couplings 
\beq
\bar{\psi}_{L}(n)i\gamma^{3}U_{L}(n, n+1)\psi_{L}(n+1), ~{\mbox{and}} ~~\bar{\psi}_{R}(n)i\gamma^{3}U_{R}(n, n+1)\psi_{R} ,
\eeq
should be separately generated for L- and R-handed branes.

As for the Higgs couplings, we have to generate 
\bea
& &q_{L}(n)^{\dagger} H_{0}(n,n) u_{R}(n)+q_{L}(n)^{\dagger} \tilde{H}_{0}(n, n) d_{R}(n) \nonumber \\
&+&q_{L}(n)^{\dagger} H_{1}(n,n+1) u_{R}(n+1)+q_{L}(n)^{\dagger} \tilde{H}_{1}(n, n+1) d_{R}(n+1),
\eea
where the first two terms connecting the same n-th branes are Wilson terms~\cite{Wilson}.
These two kinds of Higgs links connect L-handed and R-handed branes. Here $H_{i}=(\phi^{0}_{i},  \phi^{-}_{i})^{T}$ and $\tilde{H}_{i}=(-\phi^{+}_{i},
\phi_{i}^{0}*)^{T}~(i=0, 1)$ are the usual Higgs doublets, and the vacuum expectation values $\langle \phi^{0}_{i} \rangle = v_{i}~(i=1, 2)$, give the masses of the $u$ and $d$ quarks as
\begin{equation}
m_{u,d} 
=\frac{v_{0}-v_{1}\cos(p_{3}a)}{K_{u, d}}.
\end{equation}
where the hopping parameters $K_{u, d}$ are the inverse of Yukawa couplings.
If the following inequality holds
\beq
0 \le v_{0}-v_{1} \ll v_{0}+v_{1},
\label{inequality}
\eeq
then we have no doubling problem.
 Now, we may understant that to obtain the standard model, we have to prepare different branes (lattices) for L- and R-parts, and two kinds of Higgs links, $H_0$ and  $H_1$, should be generated. Gauge field on the discrete points is the theme of Alain Conne's non-commutative geometry~\cite{Conne}.  This viewpoint in our problem is clarified by Mohsen Alishahiha~\cite{Alishahiha}.
\section{Test of ADST scenario at LEP2}

This section is based on our paper by Gi-Chol Cho, Etsuko Izumi, and myself~\cite{CIS}.  Here the ADST scenario is tested using the LEP2 experiment of $e^+ e^- \to \gamma \gamma$ at $\sqrt{s}=189$ GeV.  We assume that the $x^3$-axis is generated at low energy. This newly generated axis can be considered to be fixed on the celestial sphere (against the cosmic microwave background radiation), or to move randomly. (See the analysis by Kamoshita in this point~\cite{Kamoshita}.) The dispersion (energy-momentum) relation is modified accordingly.  For $\gamma$ we have 
\beq
E^2=\sum_{i=1,2} 
p_i p^i - \left(\frac{2}{a} \sin \left(
\frac{p_3 a}{2}\right) \right)^2,
\eeq
while for electron and positron, we have
\beq
E^2=\sum_{i=1,2} 
p_i p^i - \left(\frac{1}{a} \sin \left(p_3 a \right) \right)^2 + m_{e}^2.
\eeq
In the estimation of the cross section, propagators, vertices, initial flux, phase space integration, etc. are all modified.  We assume that the initial beam axis of electron and positron is tilted from the $x^3$-axis by an angle $\chi$, but the averaging over this tilt angle $\chi$ is performed in estimating the total cross section.  We cut the scattering angle $\theta_{*}$ within $|\cos\theta_{*}|<0.9$, in accordance with LEP2 experiment~\cite{LEP2}.

Now, the total cross section of $e^+ e^- \to \gamma \gamma$ in the ADST scenario is obtained:
\beq
\sigma_{\rm ADST}(e^+ e^- \to \gamma \gamma)(s)=\sigma_{\rm QED}(s) (1+0.30 sa^2+ \cdots),
\eeq
where the ordinary QED cross section reads
\beq
\frac{d\sigma_{\rm QED}}{d(\cos\theta_{*})}=
\left(\frac{\alpha^2}{s}\right) \frac{1+\cos^2\theta_{*}}{1-\cos^2\theta_{*}},
\eeq
and
\bea
\sigma_{\rm QED} &=& 4.09 \frac{\pi \alpha^2}{s}.
\eea
Since the OPAL experiment gives 5 persent error at $\sqrt{s}=189$ GeV, the lattice constant $a$ is constrained to be
\bea
0.30 sa^2 & \le & 0.05~~{\rm for}~~\sqrt{s}=189 {\rm GeV},
\nonumber \\
{\rm or}~~a & \le & (461 {\rm GeV})^{-1}. 
\label{eq:const}
\eea
The future $e^+ e^-$ linear colliders are planned to be built, and the corresponding constraints on $a$ is found to be
$a \le 1.2$ TeV for $500$ GeV collider, 
while for $1$ TeV collider, $a$ is constrained to be less than $(2.4 {\rm TeV})^{-1}$, if the the same 5 persent error is attained, or otherwise, the discrete dimension will be observed by these future linear colliders?!

\section{TeV $\gamma$ paradox from Markarian 501 and ADST scenario}

This section is based on the forthcoming paper by Gi-Chol Cho, Isamu Watanabe, and myself~\cite{CSW}.

There are two paradoxes in cosmic rays:

(1) Ultra high energy cosmic rays (UHECR) with energy greater than $10^{20}$ eV are observed by Akeno Giant Air Shower Array observatory (AGASA)~\cite{AGASA} and others.

(2) The $\gamma$ rays up to 24 TeV are observed from the active galaxy Markarian 501 by High Energy Gamma Ray Astronomy telescope (HEGRA) at La Palma~\cite{HEGRA}.

Problem is why and how these events happen, overcoming the Greisen, Zatsepin and Kuz'min (GZK) cut~\cite{GZK}.

In this talk we study the second paradox on TeV $\gamma$ rays. The source of the TeV $\gamma$ rays is Mk 501, located 150 Mpc away from us, having the red shift value of $z=0.0336$.  The high energy $\gamma$ ray with energy $E$ can interact with the low energy back ground radiation with energy $\epsilon$, and produce the electron and positron pair. This process occurs when the following inequality holds,
\beq
E\epsilon \ge m_{e}^2=0.25 \times 10^{12}~~({\rm eV})^2.
\eeq
Therefore, high energy $\gamma$ rays with energy 1, 10, 100, and $10^3$ TeV are responsible to open the threshold of $e^{+} e^{-} \to \gamma \gamma$, against the background radiation with energy, respectively, $2.5 \times 10^{-1} {\rm eV}~(1\mu m),  2.5 \times 10^{-2} {\rm eV}~(10\mu m), 2.5 \times 10^{-3} {\rm eV}~(100\mu m), {\rm and}~~2.5 \times 10^{-4} {\rm eV}~(10^3 \mu m)$.  The first two background radiations are IR background radiations (BR), while the last one corresponds to 2.5K cosmic background radiation. Therefore, our 10 or 20 TeV $\gamma$ rays can interact with the IR BR and disappear.  Recently these IR BR with wave length $\lambda=1-100~~\mu m$ are well measured by the IR spectrometers loaded on the satellite COBE.  The observed flux of IR BR seems to indicate that the mean free path of the 20 TeV $\gamma$ rays is less than 15 Mpc, one tenth of the distance from the earth to Mk 501, so that such $\gamma$ rays could not be detected on the earth, without producing any electron and positron pairs~\cite{Meyer}. Such impossibility is called GZK cut.  To overcome this GZK cut, Coleman-Glashow, H. Sato, T. Kifume, Amelino-Camelia-Piran, and others consider the modification of dispersion relations~\cite{LorentzViolation}.  Our ADST scenario gives another possibility of modifying dispersion relations by $\sin$ functions.

Let denote (energy, momentum)'s of  TeV $\gamma$, IR $\gamma$, electron and positron as $(E, p), (\epsilon, -q), (E_{2}, p_{2})$ and $(E_{3}, p_{3})$, respectively.  Then, the energy and momentum conservation laws give
\beq
E+\epsilon=E_{2}+E_{3},~{\rm and}~~ p-q=p_{2}+p_{3},
\eeq
with
\bea
&&E=\frac{2}{a}\sin(ap/2), \epsilon=\frac{2}{a}\sin(aq/2), \nonumber \\
&&E_{i}=\sqrt{\left(\frac{1}{a} \sin (ap_{i}) \right)^2+m_{e}^2}.
\eea
The threshold conditon is modified from eq. (27) to
\beq
\epsilon \cdot E \frac{1}{2}\left(1+\sqrt{1-(aE/2)^2} \right)=m_{e}^2.
\eeq
Then, we can understand that the new threshold energy $E_{\rm th}$ and the ordinary threshold energy $E_0$ satisfy $E_0 \le E_{\rm th} \le 2E_0$, and if $a=(E_0)^{-1}$, then $E_{\rm th}=2E_0$. Therefore, $E_{\rm th}$ can be raised up to 20 TeV from the ordianary value of $E_0=10$ TeV.   Then, TeV $\gamma$ paradox disappears!?  In this case, the lattice  constant $a^{-1} \sim 10$ TeV.

\section{Conclusion}
(1) Asymptotic disappearance (discretization) of space and time (ADST) scenario is shown to be possible, at least at the final stage of generating 4d from 3d, for both gauge theory and gravity, including the standard model.

(2) QED version of the ADST scenario is consistent with the LEP2 experiment on $e^{+} e^{-} \to \gamma \gamma$, if the lattice constant $a$ of the newly generated diminsion is less than $(461~~{\rm GeV})^{-1}$.

(3) If the newly generated dimension points to the active galaxy Markarian 501, and if the corresponding lattice constant $a \sim (10~~{\rm TeV})^{-1}$, then the TeV $\gamma$ paradox from Mk 501 may be solved!?

(4) In our scenario, only a set of discrete points exists in the beginning of our universe.  There is no concept of dimensionality in such a set of discrete points.  But, if a linkage of the points is generated, a continuous dimension appears.  If four kinds of linkages can be generated one by one, during the cooling down of our universe, the present four space-time dimensions are derived.  Such ADST scenario is intereting, but to observe or not to observe it is a matter of future physics.  This is the end of my talk. Thank you.
\section*{Acknowledgements}
The author is grateful to Gi-Chol Cho, Etsuko Izumi and Isamu Watanabe for their collaboration. 
He really enjoyed staying in Tomsk during this GR11 international conference.  He gives his sincere thanks to the organizers of GR11 and its supporting staff, especially to Prof. Sergei Odintsov, Prof. Epp and Rector Obukhov, for their excellent hospitality.

\end{document}